\documentclass[Information Theory,draftcls,onecolumn]{IEEEtran}
\usepackage{cite}      % Written by Donald Arseneau
\usepackage{graphicx}  % Written by David Carlisle and Sebastian Rahtz

\usepackage{subfigure} % Written by Steven Douglas Cochran

\usepackage{amsmath}   % From the American Mathematical Society
\usepackage{amssymb}
\usepackage{bm}

\newcommand{\noi}{\noindent}
\newcommand{\la}{\lambda}

\newtheorem{rem}{Remark}

\newcommand{\km}{{\rm k_{c}}}

\newcommand{\rmx}{{\rm x}}
\newcommand{\e}{{\eta_1}}

\newcommand{\jm}{{\jmath}}

\newcommand{\atan}{{\mbox{tan}^{-1}}}

\newcommand{\dtq}{\doteq}
\newcommand{\bmthe}{{\bm \theta}}
\newtheorem{theorem}{Theorem}
\newtheorem{corollary}{Corollary}
\newtheorem{proposition}{Proposition}

\begin{document}
\title{Analytic Properties and Covariance Functions
of a New Class of Generalized Gibbs Random Fields}
\author{Dionissios~T.~Hristopulos and Samuel~N.~Elogne~\IEEEmembership{}% <-this % stops a space
%\thanks{Manuscript received December 20, 2005; revised .....
 %       }% <-this % stops a space
\thanks{The authors are with the Department of Mineral Resources Engineering,
Technical University of Crete, Chania, 73100 GREECE. (e-mail:
dionisi@mred.tuc.gr, elogne@mred.tuc.gr)}}
% note the % following the last \IEEEmembership and also the first \thanks -
% instance, "A\textbf{} \textbf{}B" will typeset as "A B" not "AB". If you want
% "AB" then you have to do: "A\textbf{}\textbf{}B"
\markboth{Submitted to IEEE Trans. Inform. Theory} {Spartan Random
Fields}
% *** Note that you probably will NOT want to include the author's name in ***
% *** the headers of peer review papers.                                   ***

% If you want to put a publisher's ID mark on the page
% (can leave text blank if you just want to see how the
% text height on the first page will be reduced by IEEE)
%\pubid{0000--0000/00\$00.00~\copyright~2002 IEEE}

%\specialpapernotice{(Invited Paper)}

\maketitle

\begin{abstract}
Spartan Spatial Random Fields (SSRFs) are generalized Gibbs random fields,
equipped with a coarse-graining kernel that acts as a low-pass filter for the
fluctuations.  SSRFs are defined by means of physically motivated spatial interactions
and a small set of free parameters (interaction couplings).
This paper focuses on the FGC-SSRF model, which is defined on the Euclidean
space $\mathbb{R}^{d}$ by means of
interactions proportional to the squares of the field realizations, as well as their
gradient and curvature.
The permissibility criteria of FGC-SSRFs are extended
by considering the impact of a finite-bandwidth kernel. It is proved that the
FGC-SSRFs are almost surely differentiable in the case
of finite bandwidth. Asymptotic explicit expressions for the Spartan covariance function are
derived for $d=1$ and $d=3$; both known and new covariance functions are obtained
depending on the value of the FGC-SSRF shape parameter.
Nonlinear dependence of the covariance integral scale on the
FGC-SSRF characteristic length is established, and it is shown that the
relation becomes linear asymptotically. The results presented in this paper
are useful in random field parameter inference, as well as in
spatial interpolation of irregularly-spaced samples.
\end{abstract}

\begin{keywords}
parameter inference, Geostatistics, Gaussian, correlations.
\end{keywords}

% Will be ignored for other modes.
\IEEEpeerreviewmaketitle

\section{Introduction}
% \PARstart{A}{demo} file is ....
% \PARstart{A}{}demo file is ....
% \PARstart{T}{his demo} file is ....
%

\PARstart{S}{patial Random} Fields (SRF's) have a wide range of
applications in hydrological models \cite{gel93,kit97,rub03},
petroleum engineering \cite{hohn}, environmental data analysis
\cite{christ,smith00,kan04}, mining exploration and mineral reserves
estimation \cite{goov,arm98}, environmental health \cite{ch98},
image analysis \cite{wink,wilson}, medical image registration
\cite{alzola} and brain research \cite{sieg95,leow04,cao} among other
fields.

A spatial random field $ \{ X({\bf s},\omega) \in \mathbb{R}; \,
{\bf s} \in D(L) \subset \mathbb{R}^{d}; \omega \in \Omega \}$ is defined as a mapping
from the probability space $(\Omega,A,P)$ into the space of real
numbers so that for each fixed ${\bf s}$,
$ X({\bf s},\omega)$ is a measurable function of $\omega$ \cite[p. 3]{abra}.
$D(L)$ is the domain within which the SRF is
defined and $L$ is a characteristic domain length.
An SRF involves by definition many possible states
\cite[p. 27]{christ},\cite{yagl87}, denoted by  $\omega$.
In the following, for notational simplicity we suppress the dependence on $\omega.$
%$\Omega$ is the \textit{sample space}, i.e., the set
%of all possible \textit{states}, $A$ is the \textit{event space}
%(which must constitute an \textit{$\sigma-$algebra}), and $P(A)$ is
%a positive \textit{probability measure} $\forall A \subset \Omega $.
%The spatial domain of interest is $D(L)$, where $L$ is a length
%characteristic of the domain size (e.g., for a square lattice, $L$
%is the length of the lattice edge.)

Realization of a particular state is determined
from a joint probability density function (p.d.f.) $f_{\rm x}\left[
X({\bf s}) \right]$. The p.d.f. depends on the spatial configuration
of the field's point values. For spatial data
the term \textit{sample} refers to $N$ values $X({\bf s}_i)$ from a
particular state at the measurement locations $\{ {\bf s}_{i}, \:
i=1,\ldots,N \}$, representing a single state of the SRF (an
observed realization).

For irregularly-spaced samples, practical applications involve determining
the statistical parameters of spatial
dependence and interpolating the data on a regular grid.
An observed realization, $X^{*}({\bf s})$, can be
decomposed into a \textit{deterministic trend} $m_{\rm x}({\bf s})$,
a \textit{correlated fluctuation SRF} $X_{\lambda}({\bf s})$, and a
random noise term, $ e({\bf s}), $ i.e., $
X^{*}({\bf s})=m_{\rm x}({\bf s})+
X_{\lambda}({\bf s})+e({\bf s}).$
The trend is a non-stationary component representing
 large-scale, deterministic
variations, which presumably correspond to the ensemble average of
the SRF, i.e. $m_{\rm x}({\bf s})=E[X({\bf s})]$.

The fluctuation
corresponds to variations that involve smaller spatial
scales than the trend. It is assumed that resolvable fluctuations
exceed the spatial resolution
 $\lambda$. The component $e({\bf s}) $ represents
inherent variability below the resolution cutoff, or completely
random variability due to uncorrelated measurement errors. It will
be assumed that $ e({\bf s}) $ is \textit{statistically independent}
of the SRF $X_{\lambda}({\bf s})$, and can be treated as
\textit{Gaussian white noise}.

It is assumed in the following that the trend has been estimated and removed.
This work focuses on modeling the correlated fluctuations.

For many geostatistical applications, the fluctuation
can be viewed as a \textit{weakly stationary SRF}, or
an \textit{intrinsic random field} with second-order stationary
increments \cite[pp. 308-438]{yagl87},\cite{math73}.
An SRF is \textit{weakly stationary} if its expectation $m_{\rm
x}({\bf s})$ is independent of the location, and its covariance
function $G_{\rm x}({\bf s},{\bf s}+{\bf r})$ depends only on the
spatial lag ${\bf r}$. In the following, the term `stationary' will
refer to weak stationarity.

Furthermore, a stationary SRF is
\textit{statistically isotropic} if the covariance function depends
only on the Euclidean distance between points but not on the
direction of the lag vector, i.e., $G_{\rm x}(|{\bf r}|)$, where
$|{\bf r}|)$ is the Euclidean norm of the vector ${\bf r}$.
The isotropic assumption is not restrictive, since the anisotropic
parameters can be inferred from the data and isotropy can be
restored by rotation and rescaling transformations
\cite{dth02,dth04,gal04,dth05}.

For isotropic, short-ranged SRF's,
one can define a single integral scale \cite[p. 22]{rub03}, given by the
integral of the covariance function along any direction in space.
Since the parameters of the fluctuation SRF are determined from the
available sample by employing the \textit{ergodic hypothesis}
\cite[p. 29]{yagl87},\cite[p. 30]{lantu}, the integral scale must be considerably smaller
than the domain size $L$.

The distribution of \textit{Gibbs random fields} is expressed in
terms of an energy functional $ H[X_\lambda({\bf s});\bmthe]$, where
$\bmthe$ is a set of \textit{model parameters} as follows:
\cite[p. 51]{wink}

\begin{equation}
\label{gibbspdf} f_{\rm x} [X_\lambda({\bf s});\bmthe] = \frac
        {\exp \left\{ { - H[X_\lambda ({\bf s});\bmthe]} \right\} }
        {Z(\bmthe)}.
\end{equation}

\noindent The constant $ Z(\bmthe) $, called the \textit{partition
function} normalizes the p.d.f. and is obtained by integrating $
\exp \left\{ -H[X_\lambda({\bf s});\bmthe] \right\} $ over all the
realizations.

Gaussian SRF's used in classical geostatistics can be included in
the formalism of Gibbs SRF's if the energy functional is expressed
as follows:
\begin{equation}
\label{covenergy1} H[X({\bf s});\bmthe]   =  \frac{1}{2}
{\sum}_{i=1}^{N} {\sum}_{j=1}^{N} X_{i} \, [G]^{-1}_{ij}({\bm
\theta}) \, X_{j}
\end{equation}
where the \textit{precision matrix}, $[G]^{-1}_{ij}(\bmthe) $ is the
inverse of the covariance matrix; the latter is determined directly
from the data by fitting to parametric models. Note that in
Eq.~(\ref{covenergy1}) there is no explicit resolution scale.

The rest of this paper is structured as follows:
Section~(\ref{sec:spartan}) gives a brief overview of Spartan
spatial random fields. Section~(\ref{sec:cov}) focuses on general
(for any $d$) properties of the covariance function.
Section~(\ref{sec:derivssrf}) proves the property of sample
differentiability for a specific class of SSRF models.
Section~(\ref{sec:1d}) focuses on a one-dimensional SSRF model and
obtains explicit expressions for the variance and the integral
scale, as well as expressions for the covariance function valid in
the infinite-band limit. Section~(\ref{sec:3d}) obtains explicit
respective expressions in three dimensions. Section~(\ref{concl})
summarizes the contributions derived in this paper. Finally, the
calculations used in deriving the variance (finite-band case) and
the covariance functions (infinite-band limit) are presented in
detail in Appendices.

\section{Overview of Spartan Spatial Random Fields}
\label{sec:spartan}

The term \textit{Spartan} indicates parametrically compact models
that involve a small number of parameters. The Gibbs property stems
from the fact that the
 \textit{joint probability
density function} (p.d.f.) is expressed in terms of an
\textit{energy functional} $H[X({\bf s})]$, i.e., $f_{\rm x}[X({\bf
s})] = \exp\{ -H[X({\bf s})] \}$. Use of an energy functional
containing terms with a clear physical interpretation permits
inference of the model parameters based on matching respective
sample constraints with their ensemble values \cite{dth03}. Thus,
the spatial continuity properties can be determined without recourse
to the variogram function.

Estimation of the experimental variogram, using the classical method
of moments or the robust estimator,  \cite{cress}, involves various
empirical assumptions (such as choice of lag classes, minimum number
of pairs per class, lag and angle tolerance, etc.
\cite[pp. 75-123]{goov},\cite[pp. 44-65]{wack}).
 In addition, inference of the `optimal' theoretical model from the
experimental variogram presents considerable uncertainties. For
example, least-squares fitting may lead to sub-optimal models. This
is due to the proportionally larger influence of larger lag
distances that correspond to less correlated fluctuations. This
situation often forces practitioners to search for a visually
optimal fit \cite[pp. 48-49]{wack} of the experimental variogram with a model
that yields better agreement in the short distance regime. It may be
possible to address some of these shortcomings more effectively in
the SSRF framework.

The recently proposed FGC-SSRF models \cite{dth03} belong in the
class of Gaussian Gibbs Markov random fields. The Markov property
stems from the short range of the interactions in the energy
functional. However, the development of SSRFs does not follow the
general formalism of GMRF's \cite{cress,besag,wink,rue,rue02}. The
spatial structure of the SSRFs is determined from the energy
functional, instead of using a transition matrix, or the conditional
probability (at one site given the values of its neighbors). Model
parameter inference focuses on determining the coupling strengths in
the energy functional, instead of the transition matrix. The GMRF
formalism and related Markov chain Monte Carlo methods can prove
useful in conditional simulations of SSRFs. Methods for the
non-constrained simulation of SSRFs on square grids and
two-dimensional irregular meshes are presented in
\cite{dth03b,dth06}. In fact, SSRF models without the Gaussian or
Markov properties can be constructed.

The energy functional of SSRFs involves derivatives (in the
continuum), suitably defined differences or kernel averages (on
regular lattices and irregular networks) of the sample
\cite{sedth06}. In all cases, the energy terms correspond to
identifiable properties (i.e., gradients, curvature). As we show
below, in the continuum case the energy functional is properly
defined only if $X_\lambda({\bf s})$ involves an intrinsic
resolution parameter `$ \lambda $'. This parameter is introduced by
means of a coarse-graining kernel. Hence, formally, the SSRFs belong
in the class of \textit{generalized random fields} \cite[p. 44]{adler},
\cite[pp. 431-446]{yagl87}.

The resolution limit $ \lambda $ is a meaningful parameter, since a
spatial model can not be expected to hold at every length scale;
also, in practice very small length scales can not be probed. In
contrast, classical SRF representations do not incorporate a similar
parameter. The resolution limit implies that the covariance spectral
density acts as a low-pass filter for the fluctuations.

For lattice-based numerical simulations, the lattice spacing
provides a lower bound for $\lambda $. The latter, in frequency
space corresponds to the upper edge, $1/2a$, of the Nyquist band;
equivalently, in wavevector space it corresponds to the upper edge,
$\pi/a$, of the first Brillouin zone. For irregularly spaced samples
there is no obvious cutoff  {\it a priori}; hence the cutoff is
treated an a model parameter to be determined from the data.
Consequently, we use the spectral band cutoff, $\km$, as the SSRF
model parameter instead of $\lambda$.

\subsection{The FGC Energy Functional }
\label{ssec:fgc_cont}

A specific type of SSRF, the fluctuation-gradient-curvature (FGC)
model was introduced and studied in \cite{dth03}. The FGC energy
functional involves three energy terms that measure the square of
the fluctuations, as well as their gradient and curvature.

The p.d.f. of the FGC model involves the \textit{parameter set}
$\bmthe=(\eta_0, \eta_1, \xi, \km)$: the scale coefficient $\eta_0$
determines the variance, the shape coefficient $\eta_1$, determines
the shape of the covariance function, and the characteristic length
$\xi$ is linked to the range of spatial dependence; the wavevector
$\km$ determines the bandwidth of the covariance spectral density.
If the latter is band-limited, $\km$ represents the band cutoff and
is related to the resolution scale by means of $\km \lambda \approx
1.$

More precisely, the FGC p.d.f. in $ \mathbb{R}^d $ is determined
from the equation:
\begin{equation}
\label{fgc} H_{\rm fgc}[X_\lambda ({\bf s});\bmthe ] =
\frac{1}{{2\eta _0 \xi ^d }} \int d{\bf s} \, \textsl{h}_{\rm fgc}
\left[ {X_\lambda({\bf s});\bmthe' } \right],
\end{equation}
where  $\bmthe'=(\eta _1 ,\xi, \km)$ is the reduced parameter set,
and $ \textsl{h}_{\rm fgc}$ is the normalized (to $\eta_0=1$) local
energy at $\bf{s}$. The functional $\textsl{h}_{\rm fgc} \left[
{X_\lambda({\bf s});\bmthe'} \right] $
  is given by the following expression

\begin{eqnarray}
\label{fgccont} h_{{\rm fgc}} \left[ {X_\lambda ({\bf s});\bmthe' }
\right] & = & \left[ {X_\lambda ({\bf s})} \right]^2  + \eta _1
\,\xi ^2 \left[ {\nabla X_\lambda ({\bf s})} \right]^2  + \xi^4
\left[ {\nabla ^2 X_\lambda ({\bf s})} \right]^2 .
\end{eqnarray}

The explicit, non-linear dependence of Eqs.~(\ref{fgc})
and~(\ref{fgccont}) on the parameters $\eta_0, \eta_1, \xi, \km$
that have an identifiable physical meaning is preferable to the use
of linear coefficients for the variance, gradient and curvature,
because it simplifies the parameter inference problem and allows
intuitive initial guesses for the parameters.

\section{The FGC Covariance Function}
\label{sec:cov}
 The FGC energy functional has a
 particularly simple expression in Fourier space.
Let the Fourier transform of the covariance function in wavevector
space be defined by means of
\begin{equation}
    \label{eq:covft0}
    \tilde{G}_{{\rm x};\lambda}({\bf k};\bmthe)= \int d{\bf r}\; e^{-\jmath \bf{k\cdot r}}
    G_{{\rm x};\lambda}({\bf r};\bmthe),
\end{equation}
and the inverse Fourier transform by means of the integral
\begin{equation}
    \label{eq:invcovft}
    G_{{\rm x};\lambda}({\bf r};\bmthe)= \frac{1}{(2\,\pi)^{d}}\,\int d{\bf k}\; e^{\jmath \bf{k\cdot r}}
    \tilde{G}_{{\rm x};\lambda}({\bf k};\bmthe).
\end{equation}
In the following, we suppress the dependence on $\bmthe$ when
economy of space requires it.

The energy functional in Fourier space is given by

\begin{equation}
\label{enfunc} H_{\rm fgc}[X_\lambda ({\bf s}) ;\bmthe ]  =
 \frac{\int d{\bf k}}{2 (2 \pi)^{d}}
\tilde{X}_{\lambda}({\bf k}) \, [\tilde{G}_{{\rm
x};\lambda}]^{-1}({\bf k}) \, \tilde{X}_{\lambda}(-{\bf k}).
\end{equation}
Note that the interaction is diagonal in Fourier space, i.e., the
\textit{precision matrix} $[\tilde{G}_{{\rm x};\lambda}]^{-1}({\bf
k})$ couples only components with the same wavevector value.

Also, for a real-valued SSRF $X({\bf s})$ it follows that
$\tilde{X}_{\lambda}(-{\bf k})=\tilde{X}^{\dagger}_{\lambda}({\bf
k})$. For a non-negative covariance spectral density, it follows
from~(\ref{enfunc}) that the energy is also a non-negative
functional.

The covariance spectral density follows from the explicit
expression:

\begin{equation}
\label{covspd} \tilde G_{{\rm x};\lambda}({\bf k}) = f_{\rm x}({\bf
k};\bmthe^{''})\, \left|\tilde Q_\lambda({\bf k})\right|^{2},
\end{equation}
where  $\bmthe^{''}=(\eta_0,\eta _1 ,\xi)$, $ \tilde Q_\lambda
({\bf{k}})\ $ is the Fourier transform of the coarse-graining kernel
and
\begin{equation}
\label{f-cov} f_{\rm x}({\bf k};\bmthe^{''})=\frac{\eta _0 \,\xi ^d
}{1 + \eta _1 \,(k\, \xi )^2 + (k\, \xi )^4 }.
\end{equation}

In \cite{dth03,dth03b}, a kernel with an isotropic boxcar spectral
density, i.e, with a sharp wavevector cut-off at $\km  $, was used.
The boxcar kernel leads to a {\it band-limited covariance spectral
density} $ \tilde G_{{\rm x};\lambda }({\bf k}) $. This kernel will
be used here as well. It involves a single parameter, i.e., $\km$,
which facilitates the inference process. Nonetheless, it is not the only possibility.

For this functional to be permissible, the covariance function must
be positive definite. If $\km \xi$ is considered as practically
infinite, application of \textit{Bochner's theorem} \cite{bochner},
\cite[p. 106]{yagl87},
permissibility requires $\eta _1> -2$, as shown in \cite{dth03}. For
negative values of $\eta_1$ the spectral density develops a sharp
peak.  $ \tilde G_{{\rm x};\lambda }({\bf k}) $ tends to become
singular as $\eta _1 $ approaches the permissibility bound of $-2$.
In early investigations \cite{dth03,var05}, $\km$ was treated as an
{\it a priori} known constant so that $\km \xi >>1$. However, it is
also possible to infer the value of $\km$ from the data
\cite{sedth06}. In this case, the permissibility criterion is
modified as follows:

\vspace{8pt}

\begin{theorem}[Permissibility of FGC-SSRF]
The FGC-SSRF is permissible (i) for any $\km$ if $\e>-2$ and (ii)
for $\e<-2$, provided that $\km \xi <
\frac{1}{\sqrt{2}}\sqrt{|\e|-\Delta}$, where $\Delta =
|\e^2-4|^{\frac{1}{2}}$.
\end{theorem}

\begin{proof}
Let us assume that $D_{k} \doteq \{k \in \mathbb{R}: \tilde Q_\lambda(k) \neq 0\}.$
Based on Eq.~(\ref{f-cov}), we obtain $f_{\rm x}(k;\bmthe^{''})=\eta _0 \,\xi ^d /\Pi(k\xi),$ where
$\Pi(\rmx)=1+\e \rmx^2+\rmx^4$. Then, $\Pi(\rmx)=(\rmx^2-y_1)(\rmx^2-y_2)$, where $y_{1,2}=(-\e
\pm \Delta)/2$. Bochner's theorem requires that $\Pi(k\xi) \ge 0, \; \forall \, k \in D_k$.
The case for $\e>-2$ is proved in \cite{dth03}. For $\e<-2$ it follows that
$y_{1,2}=(|\e| \pm \Delta)/2 >0$. Hence, Bochner's theorem is satisfied if
$\forall \, k \in D_k: k\xi < \min( \sqrt{y_1}, \sqrt{y_2})=\sqrt{(|\e| - \Delta)/2}$.
\end{proof}

\begin{rem}
Bochner's theorem is also satisfied if
$\forall \, k \in D_k: k\xi > \max( \sqrt{y_1}, \sqrt{y_2})=\sqrt{(|\e| + \Delta)/2}$.
This case corresponds to a coarse-graining kernel that acts as a high-pass filter,
and is not relevant for our purposes.
\end{rem}

\vspace{8pt}

The \textit{spectral representation} of the covariance function is given by
means of the following one-dimensional integral, where
$J_{d/2-1}(r)$ is the \textit{Bessel function of the first kind of
order} $d/2-1,$

\begin{equation}
\label{eq:cov} G_{\rm x;\lambda}({\bf r}) = \frac{\eta _0\,r\, \xi
^d }{(2\pi \, r)^{d/2} } \int_{0}^{\km} dk
 \frac{k^{d / 2} J_{d/2-1}(kr)}  {{1 + \eta _1 (k\xi )^2  + (k\xi )^4 }}.
\end{equation}

\noindent In Eq.~(\ref{eq:cov}) and in the following, we take $r$
and $k$ to represent respectively the Euclidean norms of the vector
${\bf r}$ and ${\bf k}$. Only in $d=1,$ we will use $|r|$ and $|k|$
to denote the norm (absolute value).
\noindent The Bessel function can be expanded in
a series as follows \cite[p. 359]{ww02}, where $\Gamma(x)$ is the Gamma function:

\begin{equation}
\label{besselj} J_{d/2-1}(z)= \sum_{n=0}^{\infty} \frac {(-1)^{n} }
{ \, \Gamma(n+1) \Gamma (n+\frac{d}{2}) } \left( \frac{z}{2}
\right)^{2n+d/2-1}.
\end{equation}

\subsection{The Variance} \label{ssec:var}
We investigate the dependence of the variance, $\sigma_{\rm x}^{2}
\doteq G_{\rm x;\lambda}(0)$ on the SSRF parameters.
The covariance function is well behaved at zero distance, in
spite of the singular factor $r^{d/2-1}$ dividing
the integral in Eq.~(\ref{eq:cov}). This
singularity is canceled by the leading-order term of $J_{d/2-1}(kr)$
as $r\rightarrow 0$, which is given by $J_{d/2-1}(kr) \sim
(\frac{kr}{2})^{d/2-1} / \Gamma(d/2)$.
For $r=0$ only the leading-order term of the expansion
~(\ref{besselj}) contributes.  Hence, we obtain
\begin{equation*}
  \sigma_{\rm x}^{2} = \frac{\eta_0 \xi^{d}}{
  2^{(d-2)/2}\, \Gamma\left(\frac{d}{2}\right )
  \,(2\pi)^{d/2}} \,
  \int\limits_{0}^{\km} \frac{ dk\,k^{d-1}}{1+\eta_1 (k \xi)^2+
  (k\xi)^4},
\end{equation*}
and using the variable transformation $\rmx=k \xi,$ it follows that:
\begin{equation}\label{eq:vardd}
  \sigma_{\rm x}^{2} = \frac{\eta_0 }{
  2^{(d-2)/2}\; \Gamma\left(\frac{d}{2}\right )
  \,(2\pi)^{d/2}} \,
  \int\limits_{0}^{\km\,\xi}  \frac{ d\rmx\,\rmx^{d-1}
  }{1+\eta_1 \rmx^2+ \rmx^4}.
\end{equation}
 This integral can be explicitly evaluated for any $d$ as a function of $\e$ and
$\km\,\xi.$ In the infinite-band case ($\km\,\xi
\rightarrow \infty),$ the variance integral exists only for $d<4.$

\subsection{The Integral Scale} \label{ssec:inttran}
The integral scale of the covariance function for isotropic SRFs is
given by the equation:
\begin{equation}
\label{eq:intran0} I_{d}(\bmthe') \doteq \left[ \frac{\int d{\bf
r}\, G_{\rm x;\lambda}({\bf r})}{G_{\rm x;\lambda}(0)}
\right]^{\frac{1}{d}}.
\end{equation}
Using Eq.~(\ref{eq:covft0}) with $k=0$, and Eq.~(\ref{f-cov}) for the
DC component of the spectral density, the integral scale follows
from

\begin{equation}
\label{eq:intran} I_{d}(\bmthe') = \left[ \frac{\tilde{G}_{\rm
x;\lambda}(0)}{G_{\rm x;\lambda}(0)} \right]^{\frac{1}{d}}=\xi \,
\left[ \frac{\eta_0}{G_{\rm x;\lambda}(0)} \right]^{\frac{1}{d}}.
\end{equation}

\vspace{8pt}

In sections~(\ref{sec:1d}) and~(\ref{sec:3d}) we derive
explicit expressions for the variance and the integral scale
in $d=1$ and $d=3$,
and we study their dependence on $\e$ and $\km\,\xi$.
These expressions show that the integral scale
in the preasymptotic regime is a nonlinear function of the
characteristic length, in contrast with most classical covariance
models.

% An example of a double column floating figure using two subfigures.
% (The subfigure.sty package must be loaded for this to work.)
% The subfigure \label commands are set within each subfigure command, the
% \label for the overall fgure must come after \caption.
% \hfil must be used as a separator to get equal spacing
%
%\begin{figure*}
%\centerline{\subfigure[Case I]{\includegraphics[width=2.5in]{subfigcase1}
% where an .eps filename suffix will be assumed under latex,
% and a .pdf suffix will be assumed for pdflatex
%\hfil
%\subfigure[Case II]{\includegraphics[width=2.5in]{subfigcase2}
% where an .eps filename suffix will be assumed under latex,
% and a .pdf suffix will be assumed for pdflatex
%\label{fig_second_case}}}
%\caption{Simulation results}
%\label{fig_sim}
%\end{figure*}

% An example of a floating table. Note that, for IEEE style tables, the
% \caption command should come BEFORE the table. Table text will default to
% \footnotesize as IEEE normally uses this smaller font for tables.
% The \label must come after \caption as always.
%
%\begin{table}
%% increase table row spacing, adjust to taste
%\renewcommand{\arraystretch}{1.3}
%\caption{An Example of a Table}
%\label{table_example}
%\centering
%% Some packages, such as MDW tools, offer better commands for making tables
%% than the plain LaTeX2e tabular which is used here.
%\begin{tabular}{|c||c|}
%\hline
%One & Two\\
%\hline
%Three & Four\\
%\hline
%\end{tabular}
%\end{table}

\section{Existence of FGC-SSRF Derivatives}
\label{sec:derivssrf}

In this section we prove that the band-limited FGC-SSRF models have
differentiable sample paths with probability one. Conversely, for
the infinite-band case only the first derivative of the SRF exists
in $d=1$.

Many of the covariance models used in geostatistics are
non-differentiable (e.g., the exponential, spherical, and logistic
models). Notable exceptions are the Gaussian model (which leads to
very smooth SRF realizations) and the Whittle-Mat\'{e}rn class of
covariance functions \cite{cress,sem04}; the latter include a
parameter that adjusts the smoothness of the SRF. Hence,
band-limited SSRFs enlarge the class of available differentiable SRF
models.

Non-differentiable covariance models are often selected for processes
the dynamical equations of which are not fully known or can not be solved, based solely
on the goodness of their fit to the experimental
variogram. However, this does not imply
that the sampled process is inherently non-differentiable.
If most of the candidate models are
non-differentiable, or very smooth differentiable ones, it is
not surprising that the former perform better than the latter.
Differentiable SRF models with controlled roughness may
provide equally good candidates.

\subsection{Partial Derivatives in the Mean Square Sense}
\label{ssec:mss} The existence of first and second order derivatives
of $X_\lambda ({\bf s})$ in the mean square sense is necessary to
properly define the FGC-SSRF. This follows since the energy functional,
given by Eq.~(\ref{fgccont}), involves the spatial integral of the
squares of the gradient and the Laplacian. Assuming ergodicity,
these integrals can be replaced by the respective ensemble mean
multiplied by the domain volume (in $\mathbb{R}^{d}$).

For stationary Gaussian SRFs, a sufficient condition for the field partial
derivatives to exist
 in the \textit{mean square sense} \cite[p. 24]{abra} is the following:

 Let $\overrightarrow{n} = (n_1, \ldots, n_d)$ be a vector of integer values,
  such that $n_1 + \ldots + n_d= n$. The $n$th-order partial derivative
 $\partial^{n} X({\bf s}) /
\partial s_1^{n_1 } \ldots \partial s_d^{n_d } $ exists in the mean
square sense if the following derivative of the covariance function
exists \cite{adler}

\begin{equation}
\label{nthderiv}
% G_{\rm x}^{(2n)}(0) \equiv  \left. \left[ \frac{
%d^{2n} G_{\rm x} ({\bf r}) }{dr^{2n} } \right] \right|_{{\bf r} =
%{\bf 0}}.
G_{\rm x}^{(\overrightarrow{n})}(0) = (-1)^{n} \left.
{\frac{{\partial ^{(2n)} G_{\rm x} ({\bf r})}} {\partial r_1^{2n_1 }
\ldots \partial r_d^{2n_d }} } \right|_{r=0}.
\end{equation}

\vspace{0.5 cm}

\begin{theorem}[Mean-Square Differentiability]
For FGC Spartan Spatial Random Fields with
a band-limited covariance spectral density, the partial derivatives
of any integer order $n$ are well defined in the mean square sense.
\end{theorem}

\vspace{0.5 cm}

\begin{proof} The FGC SSRFs are stationary and jointly Gaussian.
Hence, the existence of the covariance partial
derivative~(\ref{nthderiv})
 needs to be proved. It suffices to prove that
 $| G_{\rm x}^{(\overrightarrow{n})}(0)|$ exists. Equivalently,
 it suffices to
 prove the convergence of the following Fourier integral:

\setlength{\arraycolsep}{2pt}
\begin{eqnarray}
\label{spdderiv1} | G_{\rm x}^{(\overrightarrow{n})}(0)| & = &  \eta
_0 \,\xi ^d \, \int \, {d\Omega_d} \, \, \int\limits_0^\infty
{dk} \,
\left|\, \tilde{Q}_\lambda(k)\, \right|^{2}  \nonumber \\
&  & \frac{ \,k_{1}^{2n_1} \, \ldots k_{d}^{2n_d} \, k^{d - 1} }{{1 +
\eta _1 \,(k \, \xi )^2 + (k \, \xi )^4 }},
\end{eqnarray}
\setlength{\arraycolsep}{5pt}

\noi where $d\Omega_d$ is the solid angle differential.
 Let $k_{i}=k\, \cos\phi_i$ and then define the following integral over
 the unit sphere:
$Z_d=\int {d\Omega_d} \cos\phi_1\ldots \cos\phi_d$. Note that
$Z_d \le S_d=  \int \, {d\Omega_d} = 2\pi ^{d/2}  / \Gamma (d/2)
$, where $S_d$ is the surface area of the unit sphere in $d$
dimensions. Then, the $| G_{\rm x}^{(\overrightarrow{n})}(0)|$ is
given by:
\begin{equation}
\label{spdderiv2} | G_{\rm x}^{(\overrightarrow{n})}(0)|= \eta _0
\,\xi ^d \, Z_d \, \, \int\limits_0^\infty {dk} \,\, \frac{{\left|
{\, \tilde Q_\lambda(k)\,} \right|^{2} \,k^{d + 2n - 1} }}{{1 + \eta
_1 \,(k \, \xi )^2 + (k \, \xi )^4 }}.
\end{equation}

If $\tilde Q_\lambda(k)$ is the boxcar kernel,
$| G_{\rm x}^{(\overrightarrow{n})}(0)|$ in Eq.~(\ref{spdderiv2}) is
expressed in terms of the following integral:

\begin{equation}
\label{g2nzero}
    | G_{\rm x}^{(\overrightarrow{n})}(0)|  =
    \eta _0 \,\xi ^{ - 2n} \, Z_d \,\, \int\limits_0^{\km \xi } {d\rmx } \,
    \frac{{\rmx ^{d + 2n - 1} }}{{1 + \eta _1 \,\rmx ^2  + \rmx ^4 }}.
\end{equation}

\noindent The integral on the right-hand side  on the
inequality~(\ref{g2nzero}) converges for all $d$ and $n$. This
establishes the sufficient condition for the existence of partial
derivatives in the mean-square sense.
\end{proof}

\vspace{0.5 cm}

\begin{rem}
The proof focused on the boxcar kernel, but the same arguments can be used
for any kernel that decays at large $k$ faster than a polynomial.
\end{rem}

If $\km \, \xi$ is fixed, the integral of Eq.~(\ref{g2nzero}) is
proportional to $\xi^{-2n}$, implying that the SSRF is smoother for
larger $\xi$. For $ \rmx >> 1 $ the integrand behaves as $\rmx
^{d + 2n - 5} $. If $\xi$ is fixed,  the contribution of the large
$\rmx$ in the integral of Eq.~(\ref{g2nzero}) scales as
$\left(\km\right)^{2n} \left(\km \xi\right)^{d-4}$. This scaling implies
that the roughness of the SSRF increases with $\km $.

\vspace{0.5 cm}

\begin{corollary}[Infinite-Band Case] For FGC SSRFs
with an infinite band, only the second-order partial
derivative of the covariance exists in $d=1$. Higher-order
derivatives do not exist even in $d=1$, and derivatives of any order
are not permissible in any $d>1$.
\end{corollary}

\vspace{0.5 cm}

\begin{proof} For the boxcar kernel with $\km \rightarrow \infty$
(i.e. in the absence of smoothing), the integral in the right-hand side of
Eq.~(\ref{g2nzero}) converges for $d+2n <4$ and diverges in all
other cases. Convergence is attained only for $d<4$ and $n=0$ or for
$d=1$ and $n=1$.  Hence, only the first-order derivative in $d=1$
exists in the mean-square sense.
\end{proof}

\begin{rem} If a kernel that behaves asymptotically as
 $\left|{\, \tilde Q_\lambda(k)\,} \right|^{2} \propto k^{-p}$
is used instead of the boxcar, the convergence condition becomes $d+2n <4+p$.
\end{rem}

\vspace{6pt}

The existence of the first-order derivative is not sufficient to
guarantee that the FGC SSRF has second-order derivatives in the mean
square sense.  Hence, a band limit is necessary to obtain
well defined second derivatives and  the energy functional of
Eqs.~(\ref{fgc}) and (\ref{fgccont}).

Below, we derive explicit asymptotic expressions for the covariance function
in $d=1,3$ that do not admit second-order derivatives. These should be viewed as limit
forms of the FGC-SSRF model when $k_c \rightarrow \infty$.
However, it is also shown that the asymptotic expressions are accurate
estimators of the covariance function for any $\km$, provided that
$\km\xi>v_d$, where $v_d$ is a dimension-dependent constant.

\subsection{Differentiability of Sample Paths}
\label{ssec:difsp}

The existence of differentiable sample paths presupposes the
existence of the partial derivatives in the mean square sense. In
addition, a constraint on the rate of increase of the negative
covariance Hessian tensor near the origin must be satisfied \cite[p.
25]{abra} to ensure the existence of derivatives with probability one.

The \textit{negative covariance Hessian tensor} is
defined as follows:

\begin{equation}
\label{covten} G_{{\rm x};ij}({\bf r})= - \frac{ \partial^{2} G_{\rm
x}({\bf r}) }{\partial r_i \partial r_j}.
\end{equation}

\noindent The sufficient condition for the existence of the
derivative $\partial_{i}
X({\bf s})$ requires that for any $0<r<1$ there exist
positive constants $c_i$ and $\epsilon_i$, such that the following
inequality is satisfied:

\begin{equation}
\label{cond-dif} G_{{\rm x};ii}(0)-G_{{\rm x};ii}({\bf r}) \le
\frac{c_i}{| \log{r} |^{1+\epsilon_i}}.
\end{equation}

\vspace{0.5 cm}

\begin{theorem}[Existence of Path Derivatives] The FGC-SSRFs with
a band-limited covariance spectral density have differentiable
sample paths.
\end{theorem}

\vspace{0.5 cm}

\begin{proof} The FGC SSRFs are jointly Gaussian, stationary and
isotropic random fields. For an isotropic SRF, the value of the
partial derivative of $G_{{\rm x}}({\bf r})$ is independent of direction. Therefore,
$G_{{\rm x};ii}({\bf r})= -\triangle G_{\rm x}({\bf r})/d$, where
$\triangle G_{\rm x}({\bf r})= -{\sum}_{i=1}^{d} G_{{\rm x};ii}({\bf
r})$ is the Laplacian. Hence, it is sufficient to prove the validity
of the inequality~(\ref{cond-dif}) for the Laplacian, i.e.,

\begin{equation}
\label{eq:suf1} -\left[ \triangle G_{\rm x}(0) - \triangle G_{\rm
x}({\bf r}) \right] \le \frac{c}{| \log{r}|^{1+\epsilon}}.
\end{equation}

Let us define the following function:

\begin{equation}
\label{variograd} \zeta_{\rm x}({\bf r}) \dtq -\left[ \triangle
G_{\rm x}(0) - \triangle G_{\rm x}({\bf r}) \right].
\end{equation}

\noindent In light of $\zeta_{\rm x}({\bf r})$, the
\textit{sufficient condition}~(\ref{eq:suf1}) becomes:

\begin{equation}
\label{sufcon} \zeta_{\rm x}({\bf r}) \le \frac{c}{| \log{r}
|^{1+\epsilon}}.
\end{equation}

\noindent For $r \rightarrow 0$ the right hand side in the
inequality~(\ref{sufcon}) tends to zero. Hence,  the sufficient
condition requires $\zeta_{\rm x}(0) \le 0$. This is satisfied since
$\zeta_{\rm x}(0)=0$
 as it follows from the definition~(\ref{variograd}).

For $0<r<1$, the condition can be expressed as

 \begin{equation}
 \label{sufcon2}
 \zeta_{\rm x}({\bf r}) | \log{r} |^{1+\epsilon} \le c.
 \end{equation}

\noindent The Laplacian of the covariance function is given by the
following integral in wavevector space:

\begin{equation}
\label{cov:der2} -\triangle G_{\rm x}({\bf r}) = \frac{\eta _0\, \xi
^d } {(2\pi )^{d/2}\, r^{d/2-1}} \int\limits_0^{\km} {dk}
 \frac{k^{d/2 + 2} J_{d/2-1}(kr)}  {{1 + \eta _1 (k\xi )^2  + (k\xi )^4 }}.
\end{equation}

\noindent The Bessel function in Eq.~(\ref{cov:der2}) is expanded
using the series~(\ref{besselj}). The $n$th-order term in the expansion is $\propto
(k\,r)^{2n+d/2-1}$ , thus canceling the  $r^{d/2-1}$ dependence in the
denominator of $-\triangle G_{\rm x}({\bf r})$. The
leading $(n=0)$ term of $-\triangle G_{\rm x}({\bf r})$ is independent of $r$, while all other terms
vanish at $r=0$. Hence, in Eq.~(\ref{variograd}) $\triangle G_{\rm
x}(0)$ cancels the $n=0$ term of $ \triangle G_{\rm x}({\bf r})$.
The following series expansion is obtained for $\zeta_{\rm
x}({\bf r})$, in view of Eqs.~(\ref{variograd}), (\ref{cov:der2}), and (\ref{besselj}):

\begin{eqnarray}
\label{gc2} \zeta_{\rm x}({\bf r}) & = & \frac{-\eta_0 \,
\xi^{d}}{(2\pi)^{d/2}} \sum_{n=1}^{\infty}  \, \frac{(-1)^{n}\,
r^{2n}}{\Gamma(n+1) \Gamma(n+\frac{d}{2})}
\nonumber \\
& & \quad \int\limits_{0}^{\km } dk
 \frac{k^{ 2n+d+1} }{1 + \eta _1 (k\xi )^2  + (k\xi )^4 }.
\end{eqnarray}

\noindent
%===============================================================

In light of Eq.~(\ref{gc2}), the sufficient
condition~(\ref{sufcon2}) is equivalent to the following:

\begin{equation}
\label{sufcon:alt} \sum_{n=1}^{\infty} (-1)^{n+1} u_n(r) \le c,
\end{equation}

\noindent where $u_{n}(r)$ are non-negative functions given by
\begin{equation}
\label{un} u_n(r)=\frac{r^{2n}}{(2\pi)^{d/2}} \frac{| \log{r}
|^{1+\epsilon}\, \, }{\Gamma(n+1) \Gamma(n+\frac{d}{2})} A_n(\bmthe)
\end{equation}

\noindent and $A_n (\bmthe)$ represents the following integral:

\begin{equation}
\label{An} A_n (\bmthe)= \int\limits_{0}^{\km } dk\,
 k^{ 2n+d+1} \, f(k;\bmthe''),
\end{equation}

\noindent and $f(k;\bmthe'')$ is given by Eq.~(\ref{f-cov}). The
condition~(\ref{sufcon:alt}) is satisfied if the  alternating series
$\sum (-1)^{n+1} u_{n}(r)$ converges.

An alternating series converges if it is {\it absolutely convergent}
\cite[p. 18]{ww02}. According to the \textit{comparison test}
\cite[p. 20]{ww02}, the series is absolutely convergent if $u_n< C
\tilde{u}_n$, where $\sum \tilde{u}_n$ is a convergent series, and
$C$ is a constant independent of $n$.

Using the mean value theorem \cite[p. 65]{ww02}, the integral $A_n
(\bmthe)$ is evaluated as follows:

\setlength{\arraycolsep}{2pt}
\begin{eqnarray}
\label{An2} A_n (\bmthe) &=&   f(k_n;\bmthe'')
\int\limits_{0}^{\km } dk \, k^{ 2n+d+1} \nonumber\\
&=& f(k_n;\bmthe'') \left( \frac{ \km^{2n+2+d}}{2n+2+d} \right),
\end{eqnarray}
\setlength{\arraycolsep}{5pt}

\noindent where $k_n \in [0,\km], \forall n$. Let us define as
$f(k^*;\bmthe'')=\overline{\lim}_{n \rightarrow \infty}
f(k_n;\bmthe'')$ the upper limit of the sequence $f(k_n;\bmthe'')$.
The upper limit exists and is a finite number, since $\forall n$,
$f(k_n;\bmthe'') \le \max \{ f(k;\bmthe''), \, k \in [0, \km] \}$.
Then, using $\alpha \doteq 2n+2+d$, the following inequality is
obtained, $\forall \: C \in \mathbb{R} : C > f(k^*;\bmthe'')$

\begin{equation}
\label{An3} A_n (\bmthe)\le f(k^*;\bmthe'') \left( \frac{
\km^{\alpha}}{\alpha} \right) < C  \left( \frac{
\km^{\alpha}}{\alpha} \right).
\end{equation}

Based on the inequality~(\ref{An3}), the sequence of absolute values,
$u_n(r)$, of the initial series is bounded by the sequence $C \, \tilde{u}_n$,
where:

\begin{equation}
\label{unub} \tilde{u}_n = \frac{\eta_0 (\km \,
\xi)^d}{(2\pi)^{d/2}}\, \frac{\km^{2} \,| \log{r} |^{1+\epsilon}\,
(\km \, r)^{2n}} {(2n+2+d)  \, \Gamma(n+1) \, \Gamma(n+\frac{d}{2})}.
\end{equation}

\noindent To determine the convergence of the series
$\sum_{n=1}^{\infty} \tilde{u}_n$ we use \textit{d' Alembert's ratio
test} \cite[p. 22]{ww02}, which states that the series converges
absolutely if there is a fixed $n_0$, such that for all $n>n_0$,
 $| \tilde{u}_{n+1}/\tilde{u}_n | < c_0$ , where $0<c_0<1$.
Based on Eq.~(\ref{unub}), the respective ratio is given by

\begin{equation}
\label{uratio} \big| \frac{\tilde{u}_{n+1}}{\tilde{u}_n} \big| =
(\km r)^2 \, \beta(n,d) ,
\end{equation}
where
\setlength{\arraycolsep}{2pt}
\begin{eqnarray}
\label{bet} \beta(n,d) & =  & \frac{ (n+1+\frac{d}{2}) \,
\Gamma(n+1) \, \Gamma(n+\frac{d}{2})  } { (n+2+\frac{d}{2}) \,
\Gamma(n+2) \, \Gamma(n+1+\frac{d}{2}) }
\nonumber \\
 &  =  &
\frac{  1 }{(n+1) \, (n+1+\frac{d}{2}) }.
\end{eqnarray}
The function $\beta(n,d)$ is monotonically decreasing with $n$. For
fixed $\km, \, r$ let us define as $n_0$ the smallest integer for
which $ \beta(n,d) \le (\km r)^{-2}$. Then,
$|\tilde{u}_{n+1} /\tilde{u}_n |<1, \: \forall \, n >n_0$.
This concludes the proof of sample path differentiability.

\end{proof}

\vspace{8pt}

\begin{rem}
Note that higher values of $\km$ lead to higher threshold values
$n_0$, implying a slower convergence of the series~(\ref{gc2}) and
thus rougher SSRFs. Hence, $\km$ provides a handle that permits
controlling the roughness of the SSRF.
\end{rem}

\vspace{0.5cm}

\section{Covariance of One-dimensional FGC-SSRF Model}
\label{sec:1d} The 1D SSRF model can be applied to the analysis of
time series and spatial data from one-dimensional samples
(e.g., from drilling wells).

Based on Eq.~(\ref{covspd}), the 1D covariance spectral density is
given by the following expression
\begin{equation}
\label{covspd1} \tilde{G}_{\rm x;\lambda}(k;\bmthe)=
 \frac{\big | \tilde Q_\lambda  (k) \big |^2  \,\eta _0 \,\xi  }{1 + \eta _1 \,
 (k \,\xi )^2  + (k\, \xi )^4 }.
\end{equation}
The covariance function is then obtained from the inverse Fourier
transform of Eq.~(\ref{eq:invcovft}), i.e.,
\begin{eqnarray*}
\label{cov1d} G_{\rm x}(r)&=&\int\limits_{-\infty}^{\infty}
\frac{d\, k }{2\pi} \,
              \tilde{G}_{\rmx}(k)\exp({\jmath \, k r})\\
              &=&
              \frac{\eta _0 \xi}{\pi}
              \int\limits_{0}^{\infty} dk
              \frac{ \big | \tilde Q_\lambda  (k)\big |^2  \cos( k\, r)  }{1 + \eta _1 \,
 (k\, \xi )^2  + (k\, \xi )^4 }.
\end{eqnarray*}
Using the change of variables ${\rmx}=k\, \xi,$ and focusing on the
boxcar kernel, we obtain
\begin{equation}
\label{cov1d2} G_{\rm x}(r)=
              \frac{\eta _0 }{\pi}
              \int\limits_{0}^{\km \xi} d{\rmx}
              \frac{ \cos( {\rmx}\xi^{-1} r)  }{1 + \eta _1 \,
 {\rmx}^2  + {\rmx}^4 }.
\end{equation}
Next, we calculate the variance and the integral scale of the
covariance function for general $\km$, and we provide explicit
asymptotic expressions for the covariance function for $\km \xi \rightarrow
\infty$. It can be shown numerically that the asymptotic expressions are
 accurate for $\km \xi>2$, except for the differentiability at the origin.

First, we define the dimensionless constants

\begin{equation}
\label{eq:bplusmins}
\beta_{1,2}\dtq \frac{|2 \mp \e|^{1/2}}{2},
\end{equation}
\begin{equation}
\label{eq:omegaplusmoins} \omega_{1,2}\dtq
\left(\frac{|\e \mp \Delta|}{2}\right )^{1/2}.
\end{equation}

\subsection{The Variance}
\label{ssec:var1d} The variance is calculated based on
Eq.~(\ref{eq:vardd}).
\begin{proposition}[The FGC-SSRF Variance]\label{dimoneprop1}
The variance is linearly proportional to $\eta_0$, i.e.,
\begin{equation}
\label{eq:var1d} \sigma_{\rm x}^{2}=\frac{\eta _0}{2\pi}V_1(\e,\km
\xi)\end{equation} where the function $V_1(\e,\rmx)$ is given by the
following expressions, depending on the value of $\eta_!$:

\begin{equation}
\label{eq:V1} \setlength{\nulldelimiterspace}{0pt}
V_1(\e,\rmx)=\left\{
\begin{IEEEeqnarraybox}[\relax][c]{l's} \frac{1}{4\,\beta_1} \ln
\left( \frac{ \rmx^2+2\,\beta_1\rmx +1}{\rmx^2-2\,\beta_1\rmx
+1}\right) +  \\
\frac{1}{2\,\beta_2}\sum_{l=\pm 1}
 \atan\left(\frac{\rmx + l\, \beta_1}{\beta_2}\right),&for $|\e|<2,$\\
\atan(\rmx)+ \frac{\rmx}{1+\rmx^2},&for $\e=2$\\
\frac{2}{\Delta}\sum_{l=1,2}\frac{(-1)^{l+1}}{\omega_l}
          \atan \left(\frac{\rmx}{\omega_l}\right), &for $\e >2.$%
\end{IEEEeqnarraybox}\right.
\end{equation}

\end{proposition}
\begin{proof} The proof is given in Appendix I. \end{proof}

%============================================
%============================================
\vspace{6pt}

\begin{figure}[htbp]
\rotatebox{0}{ \resizebox{8.5cm}{8cm}{
\includegraphics{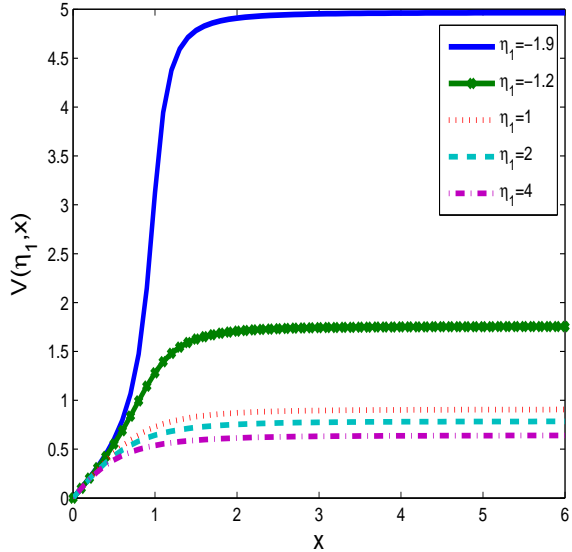}}}
\caption{Dependence of the function $V_1(\e,\rmx)$ on ${\rm x}
\equiv \km \xi$ for five different values of $\eta_1.$ }
\label{fig:V1}
\end{figure}

The variance of the Spartan model is a function of three parameters,
namely $\eta_0,$ $\eta_1$ and $\km \xi.$ Figure~\ref{fig:V1}
displays the dependence of $V_1(\e,\km \xi)$ on $\km \xi$ for
different values of the shape parameter $\eta_1$ in the range
between $-1.9$ and $4$. For fixed $\eta_1$, $V_1(\e,\km \xi)$
approaches the respective asymptotic limit for $\km \xi \geq 2.$
For fixed $\km \xi$, the
function $V_1(\e,\km \xi),$ and consequently the variance, decrease
monotonically with increasing $\eta_1$.

\subsection{The Integral Scale}
\label{ssec:ir1d} According to Eq.~(\ref{eq:intran}), the integral
scale in $d=1$ is given by
\begin{equation}
\label{eq:intran1} I_1(\bmthe')=\xi \frac{\eta_0}{G_{\rm x}(0)}=
\frac{2\pi \xi }{V_1(\eta_1,\km \xi)}.
\end{equation}
A distinct feature of the SSRF covariance functions
 is the nonlinear dependence of the
integral scale on the characteristic length $\xi$ for $\km \xi \leq
2$. However, if $\km \xi \geq 2,$ the integral scale $I_1(\bmthe')$ becomes practically
independent of the cutoff. Then, $I_1(\bmthe')$ is essentially a
function of only two variables: the shape parameter $\e$ and the
length $\xi$. The dependence on $\xi$ is linear in this
asymptotic regime. More precisely,
\begin{equation}
\setlength{\nulldelimiterspace}{0pt} I_1(\e,\xi)=\left\{
\begin{IEEEeqnarraybox}[\relax][c]{l's} 4\xi \,\beta_2,&for $|\e|<2,$\\
4\xi,&for $\e=2$\\
2\xi \, (\omega_1+\omega_2), &for $\e >2.$%
\end{IEEEeqnarraybox}\right.
\end{equation}

While the variance and the integral scale tend to asymptotic values
for $\km \xi > 2,$ this behavior does not extend to the SSRF
derivatives, i.e., to the integrals in Eq.~(\ref{g2nzero}), which
fail to converge with increasing $\km$.
 \vspace{6pt}

\subsection{Infinite-Band Covariance}
\label{ssec:infcov1} The covariance function can be evaluated
explicitly for any combination of model parameters by means of the
hyperbolic sine and cosine  functions, as well as the sine and the
cosine integrals. However, the resulting expressions are quite
lengthy. Shorter asymptotic expressions are obtained, which are valid for
$\km \xi > 2$.
More specifically:\\

\begin{proposition}[FGC-SSRF Covariance] The Spartan covariance depends linearly on the
scale factor $\eta_0$. For  $\km \xi > 2$ it  becomes a function of
the {\it normalized distance} $h \equiv |r|/\xi$ and $\e$ as
follows:
\begin{equation}
G_{\rm x}(r)=\eta _0 \, W_1(h,\eta_1)
\end{equation}
%\Big ( g_1(r) \bigone_{ |\eta_1|<
%2}+g_2(r) \bigone_{ \eta_1=2}+g_3(r) \bigone_{ \eta_1> 2}\Big )
where the function $W_1(h,\eta_1)$ is given by the following:

\vspace{6pt}

\begin{equation}
\label{eq:W1} \setlength{\nulldelimiterspace}{0pt} W_1=\left\{
\begin{IEEEeqnarraybox}[\relax][c]{l's} \, e^{-h\beta_1}\left[ \frac{\cos(h\beta_1 )}{4\,\beta_2}+
 \frac{\sin(h\beta_1)}{4\,\beta_1}
 \right],&for $|\e|<2,$\\
\frac{(1+h)}{4e^{h}},&for $\e= 2$\\
\, \frac{1}{\Delta} \, \Big(\frac{e^{-h\,\omega_{1}}}{2 \omega_{1}}-
\frac{e^{-h\,\omega_{2}}}{2 \omega_{2}}\Big), &for $\e >2.$%
\end{IEEEeqnarraybox}\right.
\end{equation}

\end{proposition}
\begin{proof}The proof is given in the Appendix II. \end{proof}

\noi \begin{corollary}[The auto-correlation function]  The
auto-correlation function is given by the equation:

\begin{equation}
\label{eq:rho1d} \setlength{\nulldelimiterspace}{0pt}
\rho(r)=\left\{
\begin{IEEEeqnarraybox}[\relax][c]{l's} \, e^{-h\beta_2}
  \left[ \cos ( h\beta_1)+\frac{\beta_2}{\beta_1}
 \sin( h\beta_1)
 \right],
 & for $|\e|<2,$\\
(1+h)\, e^{-h},&for $\e= 2$\\
\, \frac{\left( \omega_2\,e^{-h\omega_1}-
\omega_1\,e^{-h\omega_2}\right) }{ \omega_2-\omega_1}, &for $\e >2.$%
\end{IEEEeqnarraybox}\right.
\end{equation}

\begin{proof} By definition, the autocorrelation function is given by
$\rho_{\rmx}(r) \doteq G_{\rm x}(r)/\sigma_{\rm x}^{2}$.
The Eq.~(\ref{eq:rho1d}) follows from Eqs.~(\ref{eq:var1d}),
(\ref{eq:V1}), and~(\ref{eq:W1}).
\end{proof}

The autocorrelation function for $\e = 2$ corresponds to the
Whittle-Matt\'{e}rn function,
$\rho_{\nu}(r)=\frac{2^{1-\nu}}{\Gamma(\nu)}\, r^{\nu}\, K_{\nu}(r)$
with $\nu=3/2$ \cite{sem04}.  It is interesting to note that the
Whittle-Matt\'{e}rn
covariance functions are obtained by solving a stochastic equation
with an external white-noise forcing, while the Spartan covariance function
is obtained from a Gibbs energy functional. For $\e
>2$, an empirical correlation function is obtained \cite{buell}. The
correlation function for $|\e|<2$ provides a class of positive
definite functions in $d=1$, which, to our knowledge, is new.
\end{corollary}
\begin{rem}
The correlation function obtained for $|\e|<2$ in
Eq.~(\ref{eq:rho1d}) is not merely a superposition of permissible
models, since the second term contains a sine function. However, the
superposition of the two terms with the precise coefficients ensures
the permissibility and the differentiability of the correlation
function, in spite of the exponential term.
\end{rem}

\begin{figure}[htbp]
\rotatebox{0}{ \resizebox{8.5cm}{8cm}{
\includegraphics{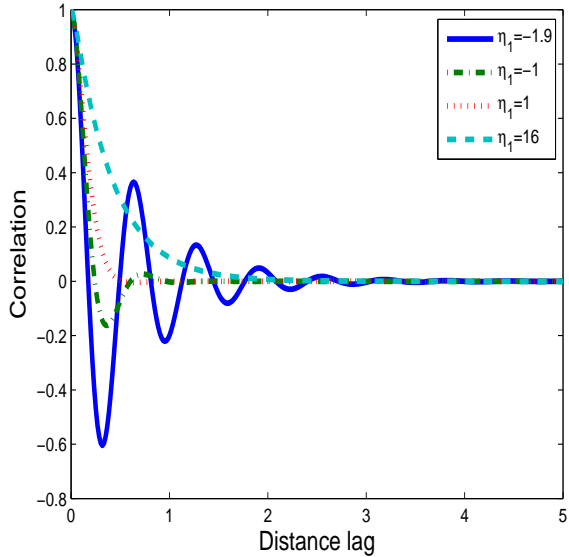}}}
\caption{Dependence of the autocorrelation function on distance for different values of
the shape parameter $\e$ with fixed $\xi=0.1.$ } \label{fig:rho1}
\end{figure}

Plots of the correlation function are shown in
Figure~(\ref{fig:rho1}) for different values of the shape parameter.
For $\e<0$ the correlation function oscillates, and
the number of oscillations increases as $\e \rightarrow -2$.
The oscillations disappear for positive
values of $\e$, and a monotonic decline of the correlations due to
the exponential terms sets in.

%===============================================================
%==============================================================

\vspace{0.5cm}

\section{Covariance of Three-dimensional FGC-SSRF Model}
\label{sec:3d}
%===========================================================

The spectral representation, i.e., Eq.~(\ref{eq:cov}), of the
isotropic Spartan covariance in $d=3$ is given by
\begin{equation}\label{eq:cov3d}
  G_{\rm x}({\bf r} ) = \frac{\eta_0 \xi^3 \,}{(2\pi)^{3/2} \,r^{1/2}} \,
  \int\limits_{0}^{\infty } dk\, \frac{ k^{3/2} \, J_{1/2}( kr)
  \, \big |\tilde{Q}_{\la}(k) \, \big
    |^2}{1+\eta_1 (k\xi)^2+ (k \xi)^4}
\end{equation}
 Using the identity
$$J_{1/2}(r)=\left (\frac{2}{\pi}\right )^{1/2}
\frac{\sin(r)}{r^{1/2}},
%\quad \mbox{or}\quad \left
%[\frac{\sin(r)}{r}= \left (\frac{\pi}{2}\right )^{1/2}
%\frac{J_{1/2}(r)}{r^{1/2}} \right ].
$$
it follows that
\begin{equation*}
  G_{\rm x}({\bf r} ) = \frac{\eta_0 \xi^3 \,}{2\pi^{2} \,r} \,
  \int\limits_{0}^{\infty } dk \,\frac{k\, \sin(k r )
  \, \big |\tilde{Q}_{\la}(k) \, \big
    |^2}{1+\eta_1 (k \xi)^2+ (k \xi)^4} .
\end{equation*}
Using the transformation $u=k \xi,$ and the boxcar kernel spectral
density, we find
\begin{equation}\label{defidim33}
  G_{\rm x}({\bf r} ) = \frac{\eta_0 \xi \,}{2\pi^{2} \,r} \,
  \int\limits_{0}^{\km \xi} du \,\frac{ u\, \sin(r  u\xi^{-1})
  }{1+\eta_1 u^2+ u^4}  .
\end{equation}
%==================================================================
\subsection{The Variance}
\label{ssec:var3d} The SSRF variance is calculated based on
Eq.~(\ref{eq:vardd}). We use the dimensionless quantities
$\beta_1$, $\beta_2$, $\omega_1$, and $\omega_2$ defined in
Section~(\ref{sec:1d}).

\begin{proposition}[FGC-SSRF Variance] The variance of
the Spartan covariance in $d=3$ is given by

\begin{equation}
\label{eq:var3d} G_{\rm x}(0)=\frac{\eta _0}{4\pi^2}V_3(\e,\km \xi)
\end{equation}

 where

 \begin{equation}
\label{eq:V3} \setlength{\nulldelimiterspace}{0pt} V_3=\left\{
\begin{IEEEeqnarraybox}[\relax][c]{l's} \frac{1}{4\,\beta_1} \ln \left( \frac{
\rmx^2-2\,\beta_1\rmx +1}{\rmx^2+2\,\beta_1\rmx +1}\right) +  \\
\frac{1}{2\,\beta_2}\sum_{l=-1,1}
 \atan\left(\frac{\rmx + l\, \beta_1}{\beta_2}\right),&for $|\e|<2,$\\
 \atan(\rmx)-
\frac{\rmx}{1+\rmx^2},&for $\e=2$\\
\frac{2}{\Delta}\sum_{l=1,2}(-1)^{l}\,\omega_l\,
              \atan \left(\frac{\rmx}{\omega_l}\right), &for $\e >2.$%
\end{IEEEeqnarraybox}\right.
\end{equation}

\end{proposition}
\begin{proof} The proof is presented in the Appendix III.
\end{proof}

\begin{figure}[htbp]
\rotatebox{0}{ \resizebox{8.5cm}{8cm}{
\includegraphics{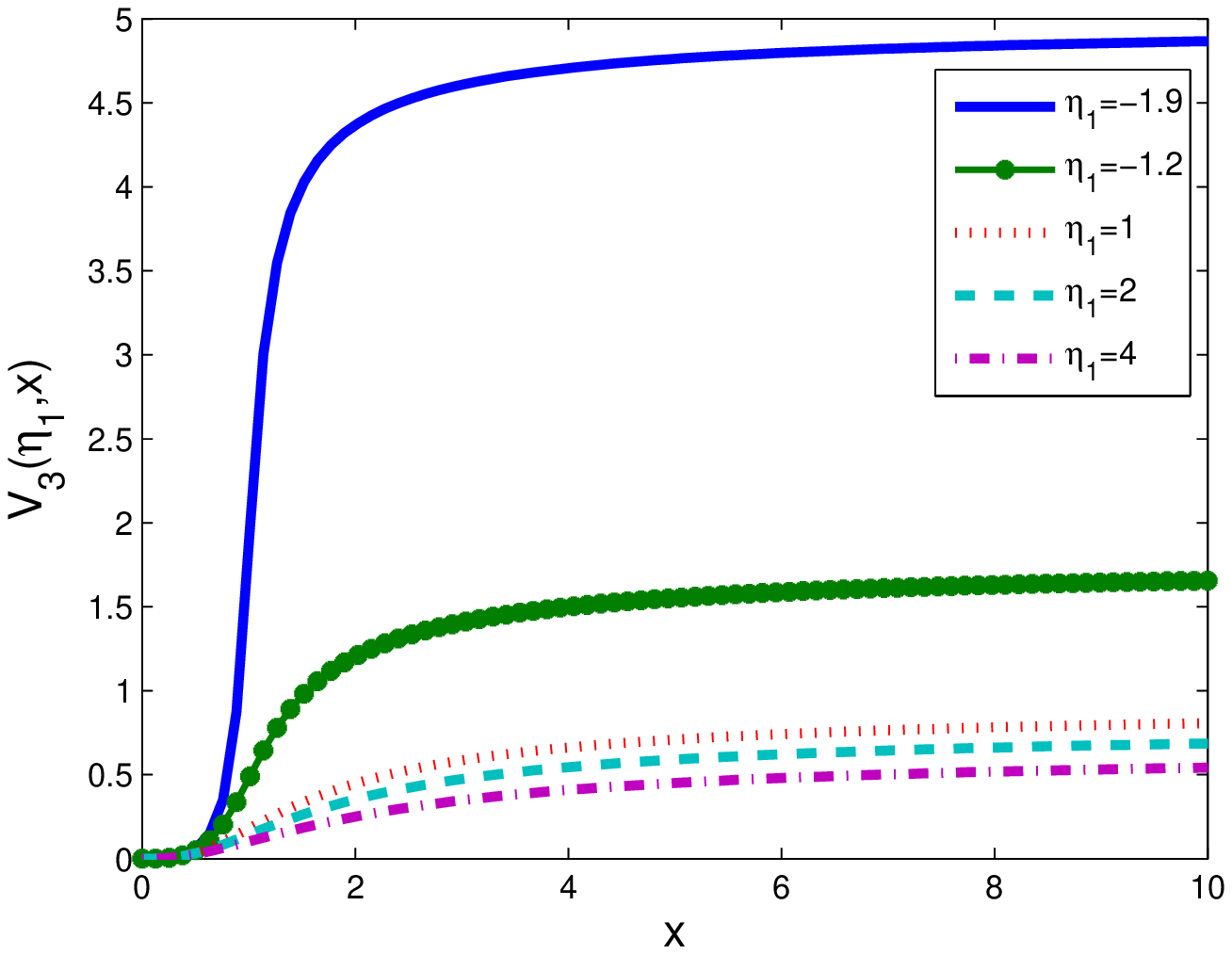}}}
\caption{Dependence of the function $V_3(\e,\rmx)$ on ${\rm x}
\equiv \km \xi$ for five different values of $\eta_1.$ }
\label{fig:V3}
\end{figure}

%=============================================================
\vspace{8pt}
%==============================================================
\subsection{The Integral Scale}
\label{ssec:ir3d} According to Eq.~(\ref{eq:intran}), the integral
scale in $d=3$ is given by
\begin{equation}
\label{eq:intran3} I_3(\bmthe')=\xi \left[\frac{\eta_0}{G_{\rm
x}(0)}\right]^{1/3}= \xi \, \left[\frac{2\pi^2 }{V_3(\eta_1,\km
\xi)}\right]^{1/3}.
\end{equation}
As seen in Figure~(\ref{fig:V3}), the approach of $V_3(\eta_1,\km
\xi)$ to the asymptotic limit is slower than in $d=1$. The
integral scale $I_3(\bmthe')$ becomes practically independent of the
cutoff for $\km \xi >5$. The asymptotic expressions
of the integral scale are:
\begin{equation}
\label{eq:intran3as} \setlength{\nulldelimiterspace}{0pt}
I_3(\e,\xi)=\left\{
\begin{IEEEeqnarraybox}[\relax][c]{l's}2\,\xi \left (\pi \,\beta_2\right)
^{1/3}&for $|\e|<2,$\\
2\,\xi \,\pi
^{1/3} &for $\e=2$\\
2\xi \left [\frac{\pi (\omega_1+\omega_2)}{2}\right]^{1/3}, &for $\e >2.$%
\end{IEEEeqnarraybox}\right.
\end{equation}

 \vspace{6pt}

%=============================================================
\vspace{8pt}
%==============================================================

\subsection{Infinite-Band Covariance}
\label{ssec:infcov3}
 As in $d=1$, the Spartan covariance function can be evaluated explicitly
for any $\bmthe$ by means of the hyperbolic
sine and cosine functions, as well as the sine and the cosine
integrals. Here we give the asymptotic (in $\km$) expressions:\\

\begin{proposition}[FGC-SSRF Covariance]
 The  covariance for $\km \rightarrow \infty$ is
expressed  as follows:
\begin{equation}
G_{\rm x}(r)=\frac{\eta_0}{2\pi} \, W_3(r/\xi,\eta_1),
\end{equation}
%\Big ( g_1(r) \bigone_{ |\eta_1|<
%2}+g_2(r) \bigone_{ \eta_1=2}+g_3(r) \bigone_{ \eta_1> 2}\Big )
where the function $W_3(h,\eta_1), \; h=r/\xi$ is given by the
following:

\vspace{6pt}

\begin{equation}
\label{eq:W3} \setlength{\nulldelimiterspace}{0pt}
W_3(h,\eta_1)=\left\{
\begin{IEEEeqnarraybox}[\relax][c]{l's} \: \frac{e^{-h\beta_1}}{\Delta}
\left[ \frac{ \sin\left(h\beta_2\right)}{h} \right],&for $|\e|<2,$\\
\: \frac{1}{4}\,e^{-h},&for $\e= 2$\\
 \: \frac{1}{2\,\Delta}\left(\frac{e^{-h\omega_1}-
e^{-h\omega_2}}{h}\right), &for $\e >2.$%
\end{IEEEeqnarraybox}\right.
\end{equation}

\vspace{8pt}

\begin{proof} The proof is given in the Appendix IV. \end{proof}

\end{proposition}

The known exponential covariance is obtained for $\e=2$,
while for $|\e|<2$ a product of two permissible models, i.e.,
$\exp(h)$ and $\sin(h)/h$, is obtained. The covariance model obtained for
$\e>2$ is new, at least to our knowledge.

\noi \begin{corollary}[The autocorrelation function] The
auto-correlation function is given by the equation:

\begin{equation}
\label{eq:rho3d} \setlength{\nulldelimiterspace}{0pt}
\rho(r)=\left\{
\begin{IEEEeqnarraybox}[\relax][c]{l's}  e^{-h\beta_1}
\left[ \frac{ \sin\left(h\beta_2\right)}{h\beta_2} \right],&for $|\e|<2,$\\
e^{-h},&for $\e= 2$\\
\frac{e^{-h\omega_1}-
e^{-h\omega_2}}{h(\omega_2-\omega_1)}, &for $\e >2.$%
\end{IEEEeqnarraybox}\right.
\end{equation}

\end{corollary}

\vspace{8pt}

\begin{proof}
It follows from the definition of the autocorrelation function, as well as
Eqs.~(\ref{eq:var3d}),
(\ref{eq:V3}), and~(\ref{eq:W3}).
\end{proof}

\vspace{8pt}

\begin{figure}[htbp]
\rotatebox{0}{ \resizebox{8.5cm}{8cm}{
\includegraphics{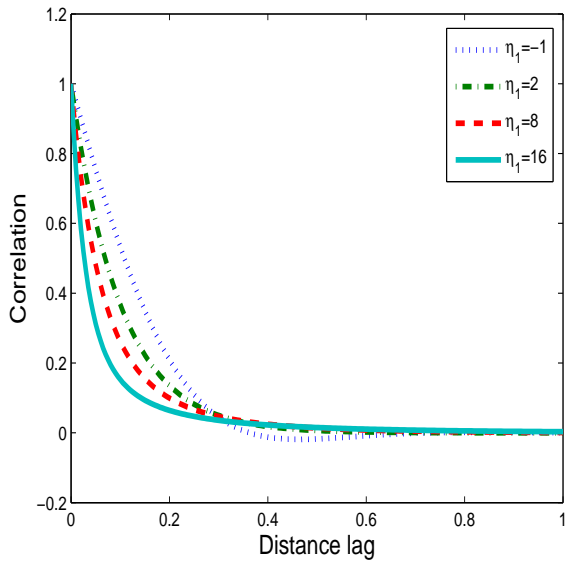}}}
\caption{Dependence of the autocorrelation function on distance for different
values of the shape parameter $\e$ with fixed $\xi=0.1$ } \label{fig:rho3}
\end{figure}

The dependence of the autocorrelation function on distance for various
values of $\e$ is shown in Figure~(\ref{fig:rho3}). Note that the
negative hole for $\e=-1$ is significantly less pronounced compared
to the $d=1$ case.

\vspace{8pt}

\begin{rem}
The Spartan covariances obtained for infinite band in $d=3$ are
continuous but
non-differentiable, in contrast with the $d=1$ case.
\end{rem}

In $d=2$ the integral of the covariance function can not be
evaluated explicitly, even in the infinite band case. However, the
covariance functions obtained for $d=3$ are also
permissible $d=1,2$ \cite[pp. 30-31]{abra}. Of course, they are not derived from
 the FGC-SSRF model in $d<3$.

%==================================================================

\section{Discussion and Conclusions}
\label{concl} We showed that FGC Spartan Spatial Random
Fields are analytic in any dimension, provided that the
covariance spectral density is band-limited. We
calculated the variance and the integral scale of the FGC-SSRF
covariance functions in one
and three dimensions. We also obtained explicit expressions for the
covariance function at the asymptotic
limit (i.e., the infinite-band limit). Depending on the value of the
shape parameter $\e$, the resulting expressions
were shown to recover known models or to yield new
covariance functions.

Explicit expressions for the covariance functions in $d=1,3$ are
also possible in the pre-asymptotic limit. However, they are not given here
since they are very
lengthy, and in practice it may be preferable to integrate
Eq.~(\ref{eq:cov}) numerically. It was also shown that the asymptotic
expressions are quite accurate for the covariance function (but not for its derivatives)
 when the product $\km\xi$ exceeds a finite,
dimensionality-dependent threshold. This result has practical applications in
SSRF parameter inference, since the procedure used involves matching
ensemble constraints, which are expressed in terms of the covariance function,
with respective sample constraints \cite{dth03,sedth06}.

Explicit expressions for the covariance function were not found
in $d=2$, where the presence of  $J_{0}(kr)$ prohibits closed form
integration in Eq.~(\ref{eq:cov}). Explicit expressions for the variance
are given in \cite{dth06}, and for the integral scale in \cite{sedth06}.

The FGC-SSRF covariance functions permit a continuous transition between smooth
(analytic) and rough (non-analytic) states, by controlling the spectral
band cutoff $\km$. This property is shared by the
Whittle-Mat\'{e}rn covariance functions.

Besides providing new covariance functions, the SSRF idea
focuses on representing spatial structure using
 energy functionals with clear physical interpretation.
 These lead to new possibilities
for parameter inference and spatial interpolation in geostatistical applications,
which are the target of continuing investigations \cite{dth03,dth06b,sedth06,var05}.

\appendices

\section{Proof of Proposition 1}
\begin{proof} We calculate the integrals required
for evaluating the variance of
the 1D SSRF. \\

\noi { Step (i)}: $|\eta_1|< 2.$ We define the characteristic polynomial
$\Pi(\rmx) \doteq \rmx^4+\e \rmx^2+1$. We expand $\Pi(x)$ as follows:
$$\Pi(\rmx) %=(\rmx^2+1)^2-(2-\e)\rmx^2
 =(\rmx^2+2\beta_1 \rmx  +1)
 (\rmx^2-2\beta_1 \rmx  +1),$$
where $\beta_1$ is given by Eq.~(\ref{eq:bplusmins}).

Using partial fraction expansion we obtain $$ \frac{1}{\Pi(\rmx)}=\frac{1}{8\beta_1 } \left
(\frac{2\rmx +2\beta_1}{\rmx^2+2\beta_1\rmx +1}-\frac{2\rmx
-2\beta_1}{\rmx^2-2\beta_1\rmx +1}\right )$$
\begin{equation}
+\frac{1}{2+\e}\left[ \frac{1}{1+(\frac{\rmx+\beta_1}{\beta_2})^2}+
\frac{1}{1+(\frac{\rmx+\beta_1}{\beta_2})^2}\right],\label{step1final8}\\
\end{equation}
where $\beta_2$ is given by Eq.~(\ref{eq:omegaplusmoins}).
Then, by direct integration the following is obtained:
$$
 \int d{\rmx}\, \frac{1}{\Pi(\rmx)}= \frac{1}{8\beta_1
}\log \left(\frac{\rmx^2+2\beta_1\rmx +1}{ \rmx^2-2\beta_1\rmx
+1}\right )$$ $$+ \frac{1}{4\beta_2} \left \{
\atan\left(\frac{\rmx+\beta_1}{\beta_2}\right)+
\atan\left(\frac{\rmx-\beta_1}{\beta_2}\right) \right \}.\\
$$
\vspace{6pt}

 \noi { Step (ii)}: $\eta_1=2.$  In this case one obtains:
\begin{eqnarray}
              \frac{ 1 }{\Pi(\rmx) }
              &=&\frac{1}{2} \;
             \frac{ 1+\rmx^2 }{(1 + {\rmx}^2)^2  }+ \frac{1}{2} \;
             \frac{ 1-\rmx^2 }{(1 + {\rmx}^2)^2  } \nonumber \\
             &=& \frac{1}{2(1 + {\rmx}^2)  }+
             \frac{ 1-\rmx^2 }{2(1 + {\rmx}^2)^2}.\label{case12}
\end{eqnarray}
It follows that
\begin{equation*}
\int d{\rmx} \,
              \frac{ 1 }{\Pi(\rmx) }
             %&=&\frac{1}{2} \int d{\rmx}
%             \frac{ 1 }{(1 + {\rmx}^2)  }+ \frac{1}{2} \int d{\rmx}
%             \frac{ 1-\rmx^2 }{(1 + {\rmx}^2)^2  }\\
             = \frac{\atan(\rmx)}{2}+ \frac{\rmx}{2(1+\rmx^2)}.
\end{equation*}

\vspace{6pt}

 \noi { Step (iii)}:  $\eta_1 > 2.$  In this case $\Pi(\rmx)$ is expressed as follows:
\begin{equation*}
 \Pi(\rmx)=\big (\rmx^2+\omega_1^2\big )\big (\rmx^2+\omega_2^2\big
),\end{equation*}
where $\omega_{1,2}$ are given by Eq.~(\ref{eq:omegaplusmoins}).
We find $$ \int d{\rmx}
              \frac{ 1 }{\Pi(\rmx) }=\frac{1}{\Delta}\int d{\rm
              x}\left
              (\frac{1}{\rmx^2+\omega_1^2}-\frac{1}{\rmx^2+\omega_2^2}\right
              )$$
              $$
              =\frac{1}{\Delta}\left [\frac{1}{\omega_1}
              \atan \left(\frac{\rmx}{\omega_1}\right
              )-\frac{1}{\omega_2}\atan
              \left(\frac{\rmx}{\omega_2}\right )\right].
$$
\end{proof}
==========================================================

\section{Proof of Proposition 2}
\begin{proof} Let us define the normalized distance $h=|r|/\xi$.
We evaluate the integrals using
integration in the complex plane. More specifically, based on the
residue theorem and \textit{Jordan's lemma} \cite[pp. 358-370]{byro92}, we obtain

\begin{equation}
   \int\limits_{0}^{\infty} d\rmx \frac{\cos(h \rmx)}{\Pi(\rmx)}=
    \Re \left\{
    \pi \jmath \, \sum \mbox{Res}^{+} \left[ \frac{\exp(\jmath h z)}{\Pi(z)} \right]
    \right\},
\end{equation}
where $\sum  \mbox{Res}^{+} \left[ Q(z) \right]$ denotes the sum of
the residues of the function $Q(z)$ in the upper half plane, and
$\Re[A(z)]$ denotes the real part of the complex function $A(z).$\\

\vspace{6pt}

 \noi { Step (i)}:  $|\eta_1| < 2.$

 Since
$\Pi(z)=\left(z^2+2\beta_1 z  +1\right)
 \left(z^2-2\beta_1 z +1\right) $
 the  imaginary poles
 in the upper half plane are
 $ z_{\pm}=\pm \beta_1+\jm\;\beta_2.$
 Thus, we obtain
 \begin{eqnarray*}& & \sum_{z=z_{\pm}} \mbox{Res}^{+}
 \left[ \frac{\exp(\jmath \, h z)}{\Pi(z)}
 \right]= \frac{\jm}{2\beta_2}\\
 && \! \! \!
 \left[ \frac{e^{-h\left(\beta_2+\jm\, \beta_1\right)
 }}{\e-2+\jm \,\Delta}+\frac{e^{-h\left(\beta_2-\jm\,
 \beta_1\right)
 }}{\e-2-\jm\,
 \Delta}\right ]
 \end{eqnarray*}
 and
\begin{eqnarray*}& &  \Re \left\{
     \jmath \sum_{z=z_{\pm}} \mbox{Res}^{+}
 \left[ \frac{\exp(\jmath \, h z)}{\Pi(z)}
 \right]\right \}= \frac{ e^{-h\beta_2}}
 {(4\e-8)\beta_2}\\
 &&
 \big[\, (\e-2)\cos\left(h\beta_1\right)
 -\Delta \sin\left(h\beta_1\right) \,
 \big]. \\
\end{eqnarray*}

\vspace{6pt}

 \noi { Step (ii)}: $\e=2$.

 Since $\Pi(z)=(1+z^2)^2$, in the upper-half plane there is a double imaginary pole,
 $z_{+}=\jmath$. Hence,
\begin{eqnarray*}
\int\limits_{0}^{\infty}
 d{\rmx}\frac{ \cos(\rmx h) }{(1 + {\rmx}^2)^2}&=&
 \Re \left\{\pi \jm
              \Big (- \jm \;\frac{1+h}{4e^{h}}\Big)
              \right \}\\
& =&   \frac{\pi (1+h)}{4e^{h}}.\\
\end{eqnarray*}

%====================================================
%=====================================================
\vspace{6pt}

 \noi { Step (iii)}:  $\eta_1 > 2.$ Using the
variables $\omega_j,$  recall from Appendix I that
$$
              \frac{ 1 }{\Pi(z) }
              =\frac{1}{\Delta}
              \left ( \frac{ 1}{z^2+\omega_1^2} -\frac{ 1}{z^2+\omega_2^2}\right).
 $$
Thus
 \begin{eqnarray*}
 \int\limits_{0}^{\infty} d{\rmx}
              \frac{ \cos(\rmx h) }{\Pi(\rmx) }&=&
              \frac{1}{\Delta}\Re \left\{\pi \jmath \,
              \Big (\frac{e^{-h\omega_1}}{2\jm \; \omega_1}
              -\frac{e^{-h\omega_2}}{2\jm \; \omega_2}\Big)
              \right \}\\
               &=&\frac{\pi}{\Delta}
              \Big (\frac{e^{-h\omega_1}}{2 \omega_1}
              -\frac{e^{-h\omega_2}}{2 \omega_2}\Big) .
 \end{eqnarray*}

 \end{proof}

\section{Proof of Proposition 3}
\begin{proof} We calculate the integrals required
for evaluating the variance of
the 3D SSRF. \\

\vspace{6pt}

 \noi { Step (i)}: $|\eta_1|< 2.$

Using the partial fraction expansion we obtain
$$\label{step1}\frac{\rmx^2}{\Pi(\rmx)}=\frac{1}{8\beta_1 } \left (\frac{2\rmx
-2\beta_1}{\rmx^2-2\beta_1\rmx +1}-\frac{2\rmx
+2\beta_1}{\rmx^2+2\beta_1\rmx +1}\right )$$ $$ +\frac{1}{2}\;\frac{
\rmx^2+1}{(\rmx^2+2\beta_1\rmx +1)(\rmx^2-2\beta_1\rmx +1)}.
$$
By direct integration we obtain
$$
 \int d{\rmx} \, \frac{\rmx^2}{\Pi(\rmx)}= \frac{1}{8\beta_1
}\log \left(\frac{\rmx^2-2\beta_1\rmx +1}{ \rmx^2+2\beta_1\rmx
+1}\right )$$ $$+ \frac{1}{4\beta_2} \left \{
\atan\left(\frac{\rmx+\beta_1}{\beta_2}\right)+
\atan\left(\frac{\rmx-\beta_1}{\beta_2}\right) \right \}.\\
$$
\vspace{6pt}

 \noi { Step (ii)}: $\eta_1=2.$  In this case the partial fraction expansion becomes

\begin{equation*}
              \frac{\rmx^2}{\Pi(\rmx)}
              = \frac{1}{2(1 + {\rmx}^2)  }-
             \frac{ 1-\rmx^2 }{2(1 + {\rmx}^2)^2}.
\end{equation*}

It follows that
\begin{eqnarray*}
\int d{\rmx} \, \frac{\rmx^2}{\Pi(\rmx)}
             %&=&\frac{1}{2} \int d{\rmx}
%             \frac{ 1 }{(1 + {\rmx}^2)  }+ \frac{1}{2} \int d{\rmx}
%             \frac{ 1-\rmx^2 }{(1 + {\rmx}^2)^2  }\\
             &=& \frac{1}{2} \left[ \, \atan(\rmx)- \frac{\rmx}{1+\rmx^2} \right].
\end{eqnarray*}

\vspace{6pt}

 \noi { Step (iii)}:  $\eta_1 > 2.$

In this case $$ \Pi(\rmx)=
\left (\rmx^2+\omega_1^2\right )\left (\rmx^2+\omega_2^2\right ),$$
and $$ \frac{\rmx^2}{\Pi(\rmx)} =\frac{1}{\Delta} \left( \frac{\omega_2^2}
{\rmx^2+\omega_2^2}-\frac{\omega_1^2}{\rmx^2+\omega_1^2} \right).$$
 It follows that

$$\int d{\rmx} \, \frac{\rmx^2}{\Pi(\rmx)}=\frac{1}{\Delta}\left [\omega_2 \,
              \atan \left(\frac{\rmx}{\omega_2} \right)
              -\omega_1 \, \atan\left(\frac{\rmx}{\omega_1}\right)\right].$$

\end{proof}

\section{Proof of Proposition 4}
\begin{proof} We follow the same procedure for integrating the covariance
spectral density as in the Appendix II.
In $d=3$ the respective integral is given by:

\begin{equation*}
 \int\limits_{0}^{\infty} d\rmx \frac{\rmx\sin(h\, \rmx)}{\Pi(\rmx)}
   =  \Im \left\{
    \pi \jmath \sum \mbox{Res}^{-} \left[ \frac{z\;e^{-\jmath \, h z}}{\Pi(z)} \right]
    \right\},
\end{equation*}
where $\sum  \mbox{Res}^{-} \left[ Q(z) \right]$ denotes the sum of
the residues of the function $Q(z)$ in the lower half plane, and
$\Im[A(z)]$ denotes the imaginary part of the complex function $A(z).$\\

 \noi { Step (i)}:  $|\eta_1| < 2.$

 The  imaginary poles
 in the lower half plane are
 $$ z_{\pm}=\pm \beta_1-\jm\;\beta_2.$$
 We have
 \begin{eqnarray*}& & \sum_{z=z_{\pm}} \mbox{Res}^{-}
 \left[ \frac{z\;e^{-\jmath \, h z}}{\Pi(z)}
 \right]= \frac{\jm}{2\beta_2}\\
 & &\left \{ \frac{\beta_1\;
 e^{-h\,\left(\beta_2+\jm \beta_1\right)
 }}{2-\e}-\frac{\beta_1\;e^{-h\,\left(\beta_2-\jm \beta_1\right)}}{2-\e}\right \}  \\
 &=& \frac{1}{\Delta} \;\sin\left (h\, \beta_1\right ) \;e^{-h\,\beta_2}.
 \end{eqnarray*}
 Thus,
\begin{equation}
   \int\limits_{0}^{\infty} d\rmx \frac{\rmx\sin(h\, \rmx)}{\Pi(\rmx)}= \frac{\pi}{\Delta}
 \;\sin\left (h\, \beta_1\right ) \;e^{-h\,\beta_2}.\\
\end{equation}
\vspace{6pt}

\noi { Step (ii)}:  $\eta_1 = 2.$
 \begin{equation*}
\int\limits_{0}^{\infty} d\rmx\;\frac{ \rmx\sin(\rmx \,h) }{\Pi(\rmx)}= \int\limits_{0}^{\infty}
 d\rmx\;\frac{ \rmx\sin(\rmx \, h) }{(\rmx^2 + 1)^2}
=  \frac{\pi \,h\,e^{-h}}{4}.
\end{equation*}

\vspace{6pt}
%====================================================
%=====================================================

\noi { Step (iii)}:  $\eta_1 > 2.$  Using
$ \Pi(z)=\left (z^2+\omega_2^2\right )\left (z^2+\omega_1^2\right )$, we obtain
$$ \frac{ z \sin(z \,h) }{\Pi(z)}
             =\frac{1}{\Delta}
              \left \{\frac{ z \sin(z \,h)}{z^2+\omega_1^2}
              -\frac{ z \sin(z \,h)}{z^2+\omega_2^2}\right\}.
 $$
 The  poles in the lower half plane are
 $ z_{+}=-\jm\,\omega_2$ and $ z_{-}=-\jm\,\omega_1.$
 It then follows that:
 \begin{eqnarray*}
&  & \int\limits_{0}^{\infty} d\rmx\;\frac{ \rmx\sin(\rmx \,h) }{\Pi(\rmx)}=
 \frac{\pi}{\Delta}\;\Im \left\{
     \jmath \; \mbox{Res}^{-} \left[ \frac{z\;e^{-\jmath \,h z}}{z^2+\omega_1^2}
     \right]_{z_-} \right. \\
    & & -\left. \jmath \; \mbox{Res}^{-} \left[ \frac{z\;e^{-\jmath \, h z}}{z^2+\omega_2^2}
    \right]_{z_+} \right\}= \frac{\pi}{2\Delta}
  \left( e^{-h\,\omega_1}-e^{-h\, \omega_2}\right).
 \end{eqnarray*}

\end{proof}

\section*{Acknowledgment}
% optional entry into table of contents (if used)
%\addcontentsline{toc}{section}{Acknowledgment}
This research is supported by the European Commission, through the
Marie Curie Action: Marie Curie Fellowship for the Transfer of
Knowledge (Project SPATSTAT, Contract No. MTKD-CT-2004-014135), and
co-funded by the European Social Fund and National Resources –
(EPEAEK-II) PYTHAGORAS.



% http://www.ctan.org/tex-archive/macros/latex/contrib/supported/IEEEtran/bibtex
%\bibliographystyle{IEEEtran.bst}

\end{document}